# The Semileptonic $\bar{B} \to D\ell\bar{\nu}$ and $\bar{B}_s \to D_s\ell\bar{\nu}$ Decays in Isgur-Wise Approach

**H. Hassanabadi*[1], S. Rahmani[1] and S. Zarrinkamar[2]**

[1] *Physics Department, Shahrood University of Technology, Shahrood, Iran*

[2] *Department of Basic Sciences, Garmsar Branch, Islamic Azad University, Garmsar, Iran*

*\* Email: h.hasanabadi@shahroodut.ac.ir*

**Abstract**

We consider a combination of linear confining and Hulthén potentials in the Hamiltonian and via the perturbation approach, report the corresponding Isgure-Wise function parameters. Next, we investigate the Isgur-Wise Function for $\bar{B} \to D\ell\bar{\nu}$ and $\bar{B}_s \to D_s\ell\bar{\nu}$ semileptonic decays and report the decay width, branching ratio and $|V_{cb}|$ CKM matrix element. A comparison with other models and experimental values is included.

## 1. Introduction

The semi-leptonic $B$ to $D$ mesonic decay is the focus of many current studies in the annals of particle physics. Although a verity of approaches have been applied to field, the relatively old but powerful Isgur-Wise function (IWF) approach can be a good candidate to analyze the problem. All form factors of semi-leptonic decays in heavy quark limit can be defined in terms of a single universal function, i.e. the IWF. The main part of the IWF includes the wave function of the meson and some kinematic factors which depend on the four velocities of heavy-light mesons before and after recoil. Decay rates, elements of CKM matrix and branching ratios can be derived from IWF [1]. There are many attempts to obtain IWF in several models [2-5]. Although different versions of the IWF exist the literature, they all assume the normalization at zero recoil, i.e. the 4-velocities ($v$ and $v'$) of mesons before and after transitions are identical. Till now, valuable papers have been released and various aspects of formalism have been discussed. Bouzas and Gupta discussed the constraints on the IWF using sum rules for $B$ meson decays [6]. Charm and bottom baryons and mesons have studied within the framework of the Bethe-Salpeter equation by Ivanov et al. and they have reported decay rates of charm and bottom baryons and mesons [7]. Kiselev determined the slope of the IWF and $|V_{cb}|$ the matrix element for semileptonic $B \to D\ell\nu$ decay [8]. The theory and phenomenology of weak decays of $B$ mesons was reviewed by Neubert [9]. Ebert et al. studied the exclusive semileptonic decays of $B$ mesons to orbitally excited $D$ mesons in the framework of the relativistic quark model [10]. Lattice study of semi-leptonic *B* Decays was well presented by Bowler et al. (UKQCD Collaboration) [11].

The mail aim of this manuscript is the study of IWF for *B* to *D* transition. In the next section we will obtain the mesonic wave function using the perturbation method. We then investigate the IWF for semileptonic *B* to *D* decay and present the slope, curvature, decay-width, branching ratio and $|V_{cb}|$ element of CKM matrix in section (3). Section (4) includes numerical results and comparison with other models. The relevant conclusions are given in section (5).

## 2. Mesonic wave function

Our starting square is the three-dimensional radial Schrödinger equation possessing the form

$$-\frac{\hbar^2}{2\mu}\nabla^2\psi_{n,\ell}(r)+V(r)\psi_{n,\ell}(r)=E_{n,\ell}\psi_{n,\ell}(r) \tag{1}$$

where $\mu$ is the reduced meson mass and $E_{n,\ell}$ denotes the energy of the system. We choose the potential as

$$V(r)=-\frac{V_0}{e^{\alpha r}-1}+br \tag{2}$$

which is a combination of a linear confinement term and the Hulthén potential. The latter behaves like a Coulomb potential for small values of *r* and decreases exponentially for large r values [12]. This behavior of the interaction in particular is interest in particle physics. Moreover, the potential has been used in other areas such as nuclear, atomic, solid-state, and chemical physics. As an example, it has been shown that the potential in the form $V_H=-\frac{V_0 e^{-\delta r}}{1-e^{-\delta r}}$, where $V_0$ and $\delta$ respectively represent the strength and screening range of the potential, can acceptably account for description of interactions between the nucleon and heavy nucleus [13]. In our calculations, we consider the linear term as the parent;

$$H_0=-\frac{\hbar^2}{2\mu}\nabla^2+br \tag{3}$$

and the Hulthén interaction therefore plays the role of the perturbation and the perturbed Hamiltonian is

$$H'=-\frac{V_0}{e^{\alpha r}-1} \tag{4}$$

As already mentioned, our parent Hamiltonian is ($\hbar=1$)

$$\frac{d^2 u_{n,\ell}}{dr^2}-\frac{\ell(\ell+1)}{r^2}u_{n,\ell}(r)-2\mu br u_{n,\ell}(r)=-2\mu E_{n,\ell}u_{n,\ell}(r) \tag{5}$$

Now let us limit the study to the ground-state with $n=1$, $\ell=0$. The corresponding equation is [14]

$$\frac{d^2 u_{1,0}}{d\kappa^2} - \kappa\, u_{1,0} = 0. \tag{6}$$

which possesses the wave function

$$u_{1,0}(r) = N\, Ai[\kappa] \tag{7}$$

where $\kappa$ defines as

$$\kappa = (2\mu b)^{\frac{1}{3}} r - (\frac{2\mu}{b^2})^{\frac{1}{3}} E_{1,0} \tag{8}$$

and $Ai$ denotes the Airy function. The corresponding energy of the system is

$$E_{1,0} = -(\frac{b^2}{2\mu})^{\frac{1}{3}} \kappa_0 \tag{9}$$

with $\kappa_0$ being the zero of Airy function which is -2.3194 in the case of ground-state ($1s$) [14]. Now, we calculate the perturbed wave function by using the first-order perturbation

$$H_0 \psi'_{1,0}(r) + H' u_{1,0}(r) = E_{1,0} \psi'_{1,0}(r) + w'\, u_{1,0}(r) \tag{10}$$

where

$$\psi'_{1,0}(r) = N'\, G(r) u_{1,0}(r) \tag{11}$$

and $w'$ is the perturbed energy. Replacing Eqs. (3), (4), (11) into Eq. (10), we can write

$$[\frac{d^2}{dr^2} + \frac{2}{r}\frac{d}{dr} - 2\mu b r + 2\mu E_{1,0}] G(r) = -2\mu w' - \frac{2\mu V_0}{e^{\alpha r} - 1}. \tag{12}$$

We now propose

$$G(z) = \sum_{m=0}^{\infty} A_m z^m \tag{13}$$

introduce the transformation $z = e^{-\alpha r}$ and use the approximation

$$\frac{1}{r^2} \approx \frac{\alpha^2}{(e^{\alpha r} - 1)^2} \tag{14}$$

to bring Eq. (12) into the form

$$\frac{d^2G}{dz^2} + \frac{1}{z}\frac{dG}{dz} - \frac{2}{(1-z)}\frac{dG}{dz} - \frac{2\mu b}{\alpha^2 z^2}\frac{(1-z)}{\alpha z}G + \frac{2\mu E_{1,0}}{\alpha^2 z^2}G = \frac{-2\mu w'}{\alpha^2 z^2} - \frac{2\mu V_0}{\alpha^2 z(1-z)} \quad (15)$$

By simplifying Eq. (15), we can write

$$\sum_{m=0}^{\infty} m(m-1)A_m z^{m-2} - \sum_{m=0}^{\infty} m(m-1)A_m z^{m-1} + \sum_{m=0}^{\infty} mA_m z^{m-2} - \sum_{m=0}^{\infty} mA_m z^{m-1} - 2\sum_{m=0}^{\infty} mA_m z^{m-1}$$
$$- \frac{2\mu b}{\alpha^3}\sum_{m=0}^{\infty} A_m z^{m-3} + \frac{4\mu b}{\alpha^3}\sum_{m=0}^{\infty} A_m z^{m-2} - \frac{2\mu b}{\alpha^3}\sum_{m=0}^{\infty} A_m z^{m-1} + \frac{2\mu E_{1,0}}{\alpha^2}\sum_{m=0}^{\infty} A_m z^{m-2} \quad (16)$$
$$- \frac{2\mu E_{1,0}}{\alpha^2}\sum_{m=0}^{\infty} A_m z^{m-1} = \frac{-2\mu w'}{\alpha^2 z^2} - \frac{2\mu V_0}{\alpha^2 z} + \frac{2\mu w'}{\alpha^2 z}$$

Equating the corresponding powers on both sides of Eq. (16), we get

$$A_0 = \frac{-w'}{E_{1,0} + \frac{2b}{\alpha}} \quad (17\text{-}a)$$

$$A_1 = -\frac{2\mu\alpha(-bw' + V_0 E\alpha + 2V_0 b)}{(E\alpha + 2b)(4\mu b + 2\mu E\alpha + \alpha^3)} \quad (17\text{-}b)$$

As a result, we have

$$\psi'_{1,0}(r) = N'(A_0 + A_1 z)u_{1,0}(r) \quad (18)$$

Thus the total wave function has the form

$$\psi^{tot}_{1,0}(r) = N^{tot}[u_{1,0}(r) + \psi'_{1,0}(r)] \quad (19)$$

where $N^{tot}$ is the normalization constant of the total wave function. We now go through the semileptonic decay $\bar{B} \to D\ell\bar{\nu}$ within the IWF approach.

## 3. Isgur-Wise Function, Decay-width and Branching Ratio of $\bar{B} \to D\ell\bar{\nu}$ decay

The IWF is often written as

$$\xi(\omega) = 1 - \rho^2(\omega - 1) + C(\omega - 1)^2 + \ldots \quad (20)$$

which is well supported by the experimental data [15]. On the other hand, the kinematic accessible region in the semileptonic decays is limited to $\omega = 1$ to 1.43. Thus, most studies consider the IWF near this region. The IWF for heavy-light mesons it can be written in terms of the integral as [16]

$$\xi(\omega) = \int_0^{\infty} 4\pi r^2 |\psi^{tot}_{1,0}(r)|^2 \cos(pr)dr \quad (21)$$

which depends on the momentum transfer ($p^2 = 2\mu^2(\omega-1)$). Because of the implication of the current conservation the form factor, the IWF is normalized to unity at $p^2 = 0$ which corresponds to $\omega = 1$ demonstrating the zero recoil limit [9]. Extending *cos(pr)* and comparing Eqs. (20) and (21), gives the so-called slope and curvature parameters as

$$\rho^2 = 4\pi\mu^2 \int_0^\infty r^4 \, |\psi_{1,0}^{tot}(r)|^2 \, dr, \tag{22-a}$$

$$C = \frac{2}{3}\pi\mu^4 \int_0^\infty r^6 \, |\psi(r)|^2 \, dr \tag{22-b}$$

where $\omega$ is referred to the mesonic zero recoil point. In Fig. (1), the behavior of IWF for some *B* and *D* mesons is plotted. The differential semileptonic decay width of $\bar{B} \to D\ell\bar{\nu}$ in the heavy-quark limit has the form [9]

$$\frac{d\Gamma(\bar{B} \to D\ell\bar{\nu})}{d\omega} = \frac{G_F^2}{48\pi^3}|V_{cb}|^2 \, (m_B + m_D)^2 m_D^3 (\omega^2 - 1)^{\frac{3}{2}} \xi^2(\omega) \tag{23}$$

where $V_{cb}$ is the element of CKM matrix. Eq. (23) indicates the dependence of the differential semileptonic decay width on the $\omega$ parameter and the product of 4-velocities of two mesons in $B \to D$ transition. We have plotted $\frac{d\Gamma}{d\omega}$ versus $\omega$ for $\bar{B} \to D\ell\bar{\nu}$ semileptonic decay in Fig. (2). By integrating of differential decay width over the interval $1 \leq \omega \leq \frac{m_B^2 + m_D^2}{2m_B m_D}$, we calculate the decay width of $\bar{B} \to D\ell\bar{\nu}$ decay. In addition, we obtain the decay width of $\bar{B}_s \to D_s\ell\bar{\nu}$ using the same approach. Fig. (3) presents the variation of $\frac{d\Gamma}{d\omega}$ vs. $\omega$ for $\bar{B}_s \to D_s\ell\bar{\nu}$ semileptonic decay. Table 1 shows our calculated slope and curvature for *D* and *B* heavy-light mesons. We have shown our results for decay width, branching ratio and the element of CKM matrix for $\bar{B} \to D\ell\bar{\nu}$ decay in Table 2. We have tabulated our mentioned results for $\bar{B}_s \to D_s\ell\bar{\nu}$ decay in Table 3. Discussions on the reported results come in the next section. As we know, the branching ratio of heavy-light meson decays obeys of

$$Br = \Gamma\tau \tag{24}$$

therefore, using the obtained decay width and the heavy-light meson life-time as $\tau_B = 1.63ps$ [17] and $\tau_{B_s} = 1.46ps$ [18], we are able to report the corresponding branching ratios as shown in third row of Tables 2 and 3.

## 4. Results and discussion

The masses of bottom and charmed B and D mesons are taken as $m_{\bar{B}} = 5.279\,GeV$, $m_D = 1.869\,GeV$, $m_{\bar{B}_s} = 5.369\,GeV$ and $m_{D_s} = 1.968\,GeV$ in the calculations [18]. We have chosen the parameters of potential as $V_0 = -1.61 GeV$, $\alpha = 0.1 GeV$, $b = 0.76 GeV^2$. The results of Table 1 are compatible with available experimental and theoretical values. Adopting the used form of IWF in the previous section, we get $\rho^2 = 1.10$ for B meson which is near the result of Skryme Model which predicted $\rho^2 = 1.3$ [5]. Sadzikowski and Zalewski reported $\rho^2_{B_s} = 1.62$ for the slope of $B_s$ meson [19]. Our result $\rho^2 = 1.75$ is in agreement with their work. Ebert et al [10] reported the decay width and the branching ratio of B to D decay as: $\Gamma = 2.7 |\frac{V_{cb}}{0.04}|^2 10^{-15} GeV$ and ( 0.63 in % ) respectively. Considering $V_{cb} = 0.04$, quantity $\Gamma$ will be 0.41 (in $10^{10}\,sec^{-1}$) [10]. Our measurements are in comparable with them. Moreover our obtained quantity for $Br(\bar{B} \to D\ell\bar{\nu}) = 1.80$ is in good agreement with reports of ARGUS, CLEO and UKQCD collaboration which are $Br(\bar{B} \to D\ell\bar{\nu}) = 2.1 \pm 0.7 \pm 0.6$ [20], $Br(\bar{B} \to D\ell\bar{\nu}) = 1.8 \pm 0.6 \pm 0.3$ [20] and $Br(\bar{B} \to D\ell\bar{\nu}) = 1.5^{+4}_{-4} \pm 0.3$, respectively [11]. We have compared our results with the reported values of decay width, branching ratio, |$V_{cb}$| for $\bar{B} \to D\ell\bar{\nu}$ decay in some other models and experimental values in the third column of Table 2.

UKQCD collaboration [11] reported $Br(\bar{B}_s \to D_s\ell\bar{\nu}) = 1.3^{+2}_{-2} \pm 0.3$. Considering $\tau_{B_s} = 1.46 ps$ [18] the value of decay width for UKQCD collaboration is $\Gamma = 0.89 \times 10^{10} s^{-1}$. Our values for this decay are in agreement with them. Moreover, our results for $\bar{B}_s \to D_s\ell\bar{\nu}$ decay are in acceptable agreement with the result of Ebert et al [10] ($\Gamma = 1.5 (in |\frac{V_{cb}}{0.04}|^2 \times 10^{-15} GeV)$, $Br\,(in\,\%) = 0.36$) which their decay width is equal to ($\Gamma = 0.22 (in\,10^{10}\,s^{-1})$). Employing the experimental decay width ($\Gamma = (1.216 \pm 0.456) \times 10^{10} s^{-1}$) for $\bar{B} \to D\ell\bar{\nu}$ [21] and using our presented model, we report CKM matrix (|$V_{cb}$|) as

$$|V_{cb}| = 0.041$$

which is acceptable when compared with available data. In the case of $\bar{B}_s \to D_s\ell\bar{\nu}$ decay we used $\Gamma = 0.89 \times 10^{10} s^{-1}$ [11].

## 5. Conclusions

We considered a mesonic system influenced by linear and Hulthén interactions. Next, using the perturbation technique, and the Isgure-Wise formalism, we obtained the corresponding decay width and branching rations for some B to D decays. Results, when compared with the exsiting data, are motivating and acceptable.

**Table 1.** slope, curvature of IWF for *B*, *D* mesons

| Meson | $\rho^2$ | $C$ |
|---|---|---|
| $B$ ($\mu_B = 0.314$) | 1.1047 | 0.3853 |
| $D$ ($\mu_D = 0.276$) | 0.9215 | 0.2693 |
| $B_s$ ($\mu_{B_s} = 0.440$) | 1.7574 | 0.9646 |
| $D_s$ ($\mu_{D_s} = 0.368$) | 1.3738 | 0.5927 |

**Table 2.** Decay width, branching ratios and $|V_{cb}|$ for $\bar{B} \to D \ell \bar{\nu}$

| Quantity | Our model | Other models |
|---|---|---|
| $\Gamma$(in $10^{10}$ sec$^{-1}$) | 1.10 | 1.413 [8] |
| | | $1.216 \pm 0.456$ [21] |
| $Br$ (in %) | 1.80 | $1.9 \pm 0.5$ [22] |
| | | 2.05 [7] |
| | | $1.79(\frac{\tau_B}{1.53ps})$ [23] |
| | | $1.7 \pm 0.4$ [24] |
| | | $1.80(\frac{\tau_B}{1.29}ps)$ [19] |
| | | $2.15 \pm 0.22$ [18] |
| $|V_{cb}|$ | 0.041 | $0.038 \pm 0.003$ [8] |
| | | 0.038 [20] |
| | | $0.042 \pm 0.001$ [25] |

**Table 3.** Decay width, branching ratios and $|V_{cb}|$ for $\bar{B}_s \to D_s \ell \bar{\nu}$

| Quantity | Our model |
|---|---|
| $\Gamma$(in $10^{10}$ sec$^{-1}$) | 0.69 |
| $Br$ (in %) | 1.00 |
| $|V_{cb}|$ | 0.045 |

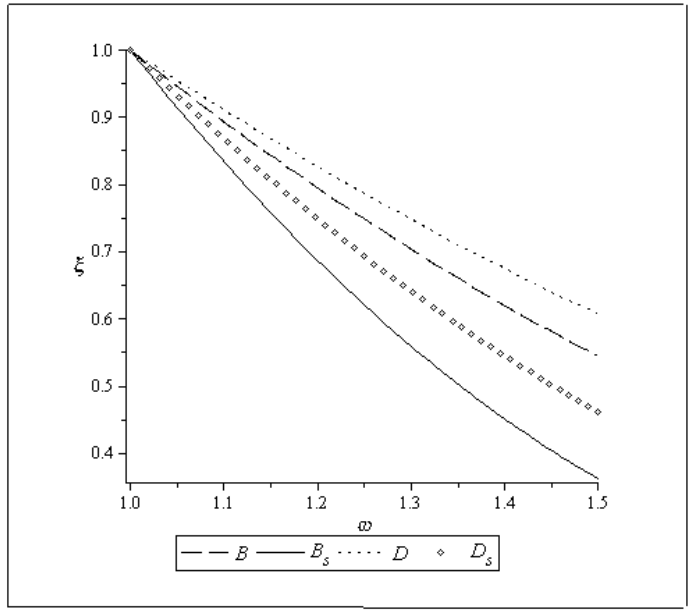

**Fig. 1.** Variation of IWF for some *B*, *D* mesons

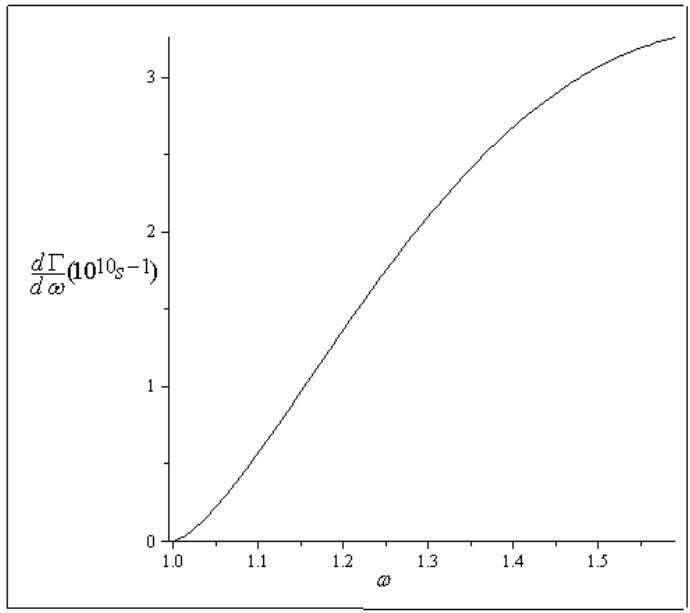

**Fig 2.** Differential decay width vs. $\omega$ for $\bar{B} \to D \ell \bar{\nu}$

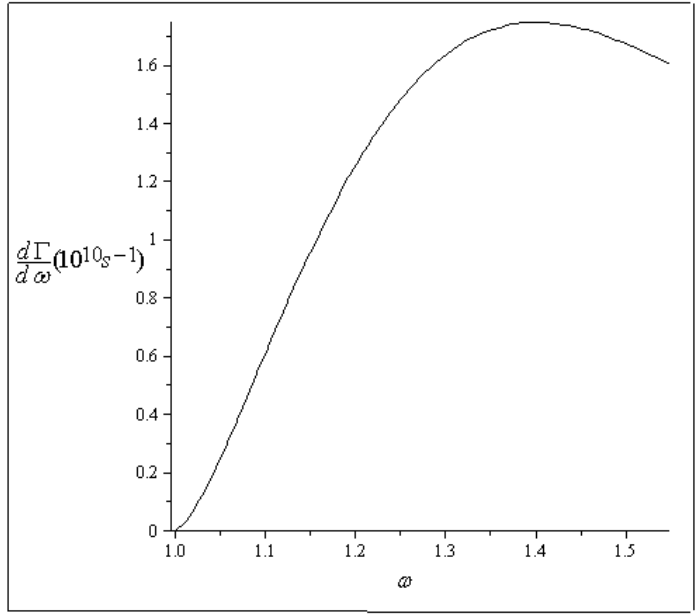

**Fig 3.** Differential decay width vs. $\omega$ for $\bar{B}_s \to D_s \ell \bar{\nu}$